\documentclass[showpacs,floatfix,aps]{revtex4}

\usepackage{amsmath,amssymb,amsfonts}
\usepackage[dvips]{graphics}

\newcommand{\be}{\begin{equation}}
\newcommand{\ee}{\end{equation}}
\newcommand{\vol}{\sqrt{g} \; dx^1\wedge\cdots\wedge dx^n}

\DeclareMathOperator{\diverg}{div}

\newtheorem{prop}{Proposition}

\begin{document}

\title{Volume elements and torsion}

\author{Ricardo A. Mosna}
\email{mosna@ime.unicamp.br}
\author{Alberto Saa}
\email{asaa@ime.unicamp.br}
\affiliation{
Instituto de Matem\'atica, Estat\'\i stica e Computa\c{c}\~ao Cient\'\i fica,
Universidade Estadual de Campinas, C.P. 6065, 13083-859,
Campinas, SP, Brazil.}


\begin{abstract}
We reexamine here the issue of consistency of minimal action formulation
with the minimal coupling procedure (MCP) in spaces with torsion. In Riemann-Cartan
spaces, it is known that a proper use of the MCP requires that the trace of
the torsion tensor be a gradient, $T_\mu=\partial_\mu\theta$,
and that the modified volume element
$\tau_\theta = e^\theta \sqrt{g} \; dx^1\wedge\cdots\wedge dx^n $ be used
in the action formulation of a physical model.
We rederive this result here under considerably
weaker assumptions, reinforcing some recent results about the inadequacy of
propagating torsion theories of gravity to explain the available observational
data.
The results presented here also open the door to possible applications of the modified
volume element in the geometric theory of crystalline defects.
\end{abstract}

\maketitle

\section{Introduction}

The use of modified volume elements in physical models has been
receiving considerable interest in recent years. In the context
of the Einstein-Cartan theory of gravity\cite{EC}, for instance,
the introduction of a new volume element gave rise to some
models possessing several interesting characteristics, such as:
propagation of torsion\cite{Saa1,Fiziev}, interaction between torsion and gauge
fields without violation of the gauge symmetry\cite{Saa2}, and a
geometrical interpretation of string-theory inspired models of
gravity\cite{SaaCQG}. Non-canonical volume elements have also proved to be
useful in other contexts. For recent applications in field
and string theory, see, for instance, \cite{Guend}.

\begin{figure}[ht]
\resizebox{0.5\linewidth}{!}{\includegraphics*{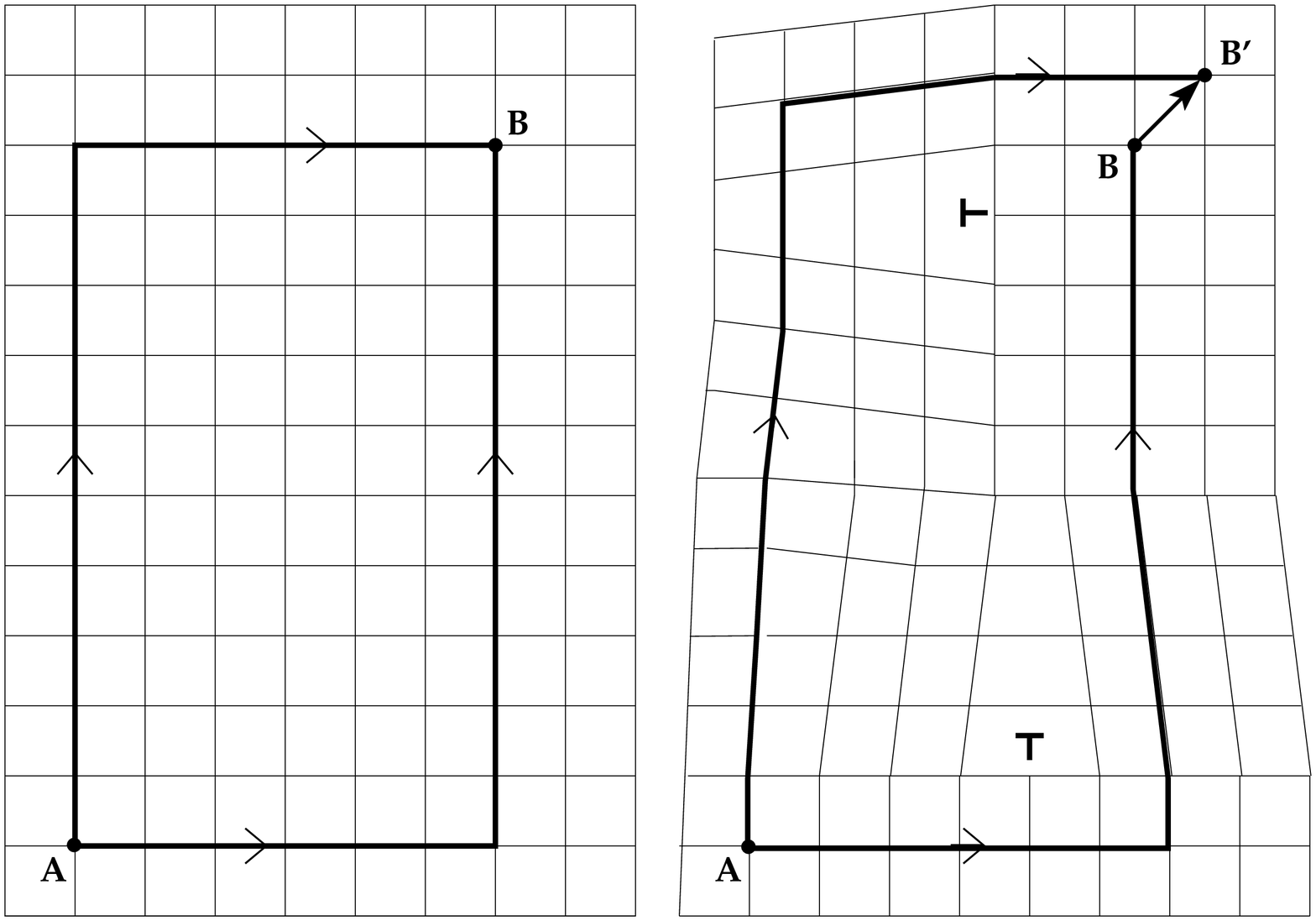}}\\
\vspace{0.5cm}
\resizebox{0.5\linewidth}{!}{\includegraphics*{torsion51.eps}}
\caption{We show, for illustrative purposes only, a lattice with two
edge dislocation defects (upper right) and a reference one  (upper left).
A closed rectangular path in the reference lattice becomes
unclosed in the presence of the defects. The vector $\overline{BB'}$,
the so-called Burgers vector, measures how the path fails to close.
In the associated geometrical description of the path
generated by the (infinitesimal) vectors $X^\mu$ and $Y^\mu$
at the point $A$,  the vector $\overline{CB}$ is obtained by parallel transport
of $Y^\mu$ along $X^\mu$, and  $\overline{DB'}$ is obtained by transporting $X^\mu$
along $Y^\mu$.  The Burgers vector in this case is given by
$\overline{BB'} =T_{\nu\kappa}{}^\mu X^\nu Y^\kappa$.
}
\label{cryst}
\end{figure}

Geometries with torsion, on the other hand, have a long tradition in
Physics. For instance, in Einstein-Cartan theory of gravity\cite{EC},
spacetime is assumed to be a Riemann-Cartan manifold, {\em i.e.}, a manifold
endowed with a Lorentzian metric and a non-symmetrical metric-compatible linear
connection. The anti-symmetrical part of the connection (the torsion tensor) is
also relevant to field theory, mainly with respect
to renormalization issues\cite{FT}. The very active area of defects in crystal
microstructures also takes advantage of these non-Riemannian geometries:
disclinations and dislocations in a crystal can be described by means of curvature
and torsion (for a revision, see \cite{Kroner,Solid}). Dislocations, in particular,
are described by the Burgers vector, which has the torsion tensor as its surface
density (see Fig.~\ref{cryst}).

The motivation for the introduction of the modified volume element proposed
in \cite{Saa3} was the observation that, in the presence of torsion,
the minimal coupling procedure (MCP) performed at the action level
is not equivalent to that performed directly at the corresponding
field equations. As shown originally in \cite{Saa3}, in Riemann-Cartan
spaces the equivalence between the MCP performed at the action and at
the field equations is recovered provided that the trace of the torsion
tensor be a gradient, $T_\mu=\partial_\mu\theta$, and that the usual volume
element $\tau_0=\sqrt{g} \; dx^1\wedge\cdots\wedge dx^n$ be deformed to
\be
\tau_\theta =e^\theta \tau_0.
\label{mve}
\ee
In \cite{Fiziev}, Fiziev reinterpreted such model as a
Transposed-Equi-Affine Theory of Gravity, since, as we will see,
the covariant derivative of densities used in \cite{Saa3} can be
related to a certain ``transposed'' connection. Such a transposed
connection, however, is a highly artificial object, being generically
not metric-compatible even when the initial connection is.
Besides, at first glance, the very motivation of demanding equivalence
between the MCP as applied to the action level and to the field equations
may be contested on physical grounds.
In fact, even in the case of General Relativity
(where the connection is torsion-free), the naive minimal coupling
between the electromagnetic field and gravity does not satisfy such
equivalence requirement if one does not choose carefully in
which set of equations the MCP should be applied.
In such a case, the field equations
contain partial derivatives of higher order, which renders
the MCP, if applied directly to some field equations, ill-defined
\cite{gravitation}. The usual solution for this kind of difficulty is to
assume  that the MCP should be always applied at the Lagrangian level.

The aim of this paper is to show that the modified volume
element (\ref{mve}) follows from a more fundamental hypothesis,
reinforcing its mathematical and physical relevance.
We will show that the existence of the modified
volume element (\ref{mve}) is a necessary and sufficient
condition to render the MCP well-defined even when considered
solely at the action level. In particular, we will show that
the results presented in \cite{Saa3} can be obtained without
recourse to transposed connections.
Our conclusions apply to any physical model where torsion is
regarded as a dynamical entity that couples to the remaining fields
via the MCP. In this way, one can expect that the results
presented here can be also useful in the geometric theory of
crystalline defects\cite{Solid,kata}, since
the expressions for the free and interaction energies of
defects and the equations for phonons and other test fields
can be obtained by arguments
very similar to the MCP.

We observe that the model proposed in \cite{Saa1} was carefully studied
in \cite{Neutron,Manoff}. By considering solutions describing compact
(neutron) stars and some solar system experiments, it was
shown\cite{Neutron} that the predictions of the model \cite{Saa1} are
in contradiction with General Relativity and, more importantly,
with the available observational data.
The propagating torsion model proposed in \cite{Saa1}, therefore, is not
a viable alternative to General Relativity.
As we will see, the results presented here make the conclusions of \cite{Neutron}
much stronger, since we will show that the existence of (\ref{mve}) is required to
ensure the very consistency of the action formulation via the MCP.
In other words, the results of \cite{Neutron} would allow us to conclude
that the very reasonable requirement of consistency of the action formulation
with the MCP {\em imply}, on account of the observational data, that the trace of
the torsion tensor of spacetime should vanish.

This work is organized as follows.
The next section is a brief review of the mathematical framework necessary to set up
the problem. Section III
contains our main results, and the last section is devoted to some final remarks.
Mathematical proofs are left to the Appendix, where we also discuss how the results
presented here relate to the earlier approach of \cite{Saa1, Fiziev, Saa2, SaaCQG, Saa3}.

\section{Notation}

In order to introduce the problem, we briefly recall some definitions.
For the sake of clarity, we employ here
a more mathematically oriented notation,
which slightly differs from that used in the original works \cite{Saa1, Fiziev, Saa2, SaaCQG, Saa3}.
Let $\cal M$ be an n-dimensional manifold.
In the following, the components of the covariant derivative of a vector field
$A=A^\nu e_\nu$, with respect to a local basis in $T\cal M$, are denoted by
$D_\mu A^\nu$, so that $\nabla_\mu A = (D_\mu A^\nu)e_\nu$, with
$D_\mu A^\nu=\partial_\mu A^\nu + \Gamma_{\mu\sigma}{}^\nu A^{\sigma}$.
The components of the torsion tensor associated with the connection $\nabla$,
\be
T(X,Y) = \nabla_X Y - \nabla_Y X - [X,Y],
\ee
are given, in a coordinate basis, by
\be
\label{torsion}
T_{\mu\nu}{}^\kappa = \Gamma_{\mu\nu}{}^\kappa - \Gamma_{\nu\mu}{}^\kappa.
\ee

Whenever the manifold $\cal M$ is endowed with a metric $g_{\mu\nu}$, the
connection coefficients $\Gamma_{\mu\nu}{}^\kappa$ can be decomposed as
\begin{equation}
\Gamma_{\mu\nu}{}^\kappa=
\genfrac{\{}{\}}{0pt}{}{\kappa}{\mu\nu}
+K_{\mu\nu}{}^{\kappa}+V_{\mu\nu}{}^{\kappa},
\label{coeficientes de conexao}
\end{equation}
where $\genfrac{\{}{\}}{0pt}{}{\kappa}{\mu\nu}$
are the Christoffel symbols associated with the underlying
Levi-Civita connection (uniquely defined metric-compatible
connection  without torsion on $({\cal M},g)$),
\be K_{\mu\nu\kappa} = -\frac{1}{2} \left( T_{\nu\kappa\mu}+
T_{\mu\kappa\nu}-T_{\mu\nu\kappa}\right)
\ee
is the contorsion tensor associated with $\nabla$, and
\be V_{\mu\nu\kappa}=-\frac{1}{2}\left(D_\mu g_{\nu\kappa}+
D_\nu g_{\kappa\mu} - D_\kappa g_{\mu\nu}\right)
\ee
provides a measure of the nonmetricity associated with $\nabla$.
Relevant quantities here are the trace of the above defined tensors and
their associated 1-forms
\begin{eqnarray}
T_\nu = T_{\mu\nu}{}^\mu, \quad t = T_\nu \, dx^\nu, \nonumber \\
V_\nu = V_{\mu\nu}{}^\mu, \quad v = V_\nu \, dx^\nu.
\label{1-forms}
\end{eqnarray}

A volume element on a $n$-dimensional manifold $\cal M$ is a nowhere vanishing
$n$-form on $\cal M$ \cite{KobayashiNomizu}. If $\cal M$ is endowed with a metric,
an arbitrary volume element $\tau$ can be written as
\be
\tau = f\tau_0,
\label{arbitrarytau}
\ee
where $f$ is a nowhere vanishing smooth function on $\cal M$ and $\tau_0$
is the canonical volume element
\be
\tau_0 = \theta^1\wedge\cdots\wedge\theta^n= \sqrt{g} \; dx^1\wedge\cdots\wedge dx^n,
\label{canonical}
\ee
where $\{\theta^{i}\}$ is a local orthonormal basis.

\section{Minimal coupling and equivalent Lagrangians}

Let $\mathcal{L}$ be a Lagrangian density describing a
given physical model defined on flat space or spacetime $\cal M$.
There is a natural equivalence relation $\sim$ on the set of such
Lagrangians, defined by $\mathcal{L}_1\sim\mathcal{L}_2$ if and only if
$\mathcal{L}_1$ and $\mathcal{L}_2$ give rise to {\em identical}
field equations on $M$.
Except for scale transformations, we then have
\begin{equation}
\mathcal{L}_1\sim\mathcal{L}_2\iff\mathcal{L}_1-\mathcal{L}_2=\diverg(X)
\label{relequiv}
\end{equation}
for some vector field $X$, where, for flat $\cal M$,
\be
\diverg(X) =\partial_\mu X^\mu.
\ee
This follows trivially from Gauss theorem, since integration of the Lagrangian density
immediately yields
\be
\int \diverg(X)\, \tau_0 = {\rm [surface\ term]},
\ee
with no contribution to the Euler-Lagrange equations. We emphasize that, in the context
of flat space (and of Special Relativity),
\emph{any choice} $\mathcal{L}$ in a given class $[\mathcal{L}]$
of Lagrangians leads to completely equivalent physical models.

Suppose now that the metric and connection on $\cal M$ become dynamical,
describing in this way physical fields interacting with
test particles or fields on $\cal M$. This would be the case, for instance, of
General Relativity, in which gravity is described by a dynamical metric
$g_{\mu\nu}$  with its associated Levi-Civita connection.
Other examples are the Einstein-Cartan theory of gravity\cite{EC}, where torsion becomes
an additional dynamical quantity, and the geometric theory of defects\cite{Kroner,Solid}, where
curvature and torsion are related to the density of dislocations and disclinations
of the crystalline lattice structure.

The minimal coupling procedure (MCP) is the standard prescription to
obtain the curved space equations from the flat space ones. It states that
one needs merely
to substitute ordinary derivatives by covariant ones and the Minkowskian
metric tensor by its non-flat counterpart. It is worth
noting that the MCP is used with success in nearly all physically relevant gauge theories,
including General Relativity \cite{fn}.

Let us denote the Lagrangian density obtained from $\mathcal{L}$, via MCP,
by $\mathcal{L}^{\ast}$. Similarly to the flat-space case discussed above,
the condition for $\mathcal{L}_{1}^{\ast}$ and $\mathcal{L}_{2}^{\ast}$ to be equivalent
(i.e., to yield the same Euler-Lagrange equations) is given by
\be
\mathcal{L}_{1}^{\ast}-\mathcal{L}_{2}^{\ast} = \diverg_\tau (X)
\label{lagrangeanas equiv esp curvo}
\ee
for some vector field $X$, with the definition of divergence being given
now by \cite{KobayashiNomizu}
\be
(\diverg_\tau X) \; \tau = \pounds_X \, \tau,
\label{def diverg}
\ee
where $\tau$ is a volume form on $\cal M$ and
$\pounds_X$ denotes the Lie derivative along the vector field $X$. This is
the most natural definition of divergence in this context, since
$\pounds_X \, \tau =  d\,i_X\tau$ (where $d$ denotes the exterior derivative and
$i_X$ the contraction by $X$ \cite{KobayashiNomizu}) leads directly to Gauss theorem
\be
\int_{\cal M} \diverg_\tau(X)\tau=\int_{\partial \cal M}i_X \tau = \textrm{[surface term]}.
\ee

If $\mathcal{L}_{1}$ and $\mathcal{L}_{2}$ are equivalent Lagrangians in flat-space,
we get, after applying the MCP,
\begin{equation}
\mathcal{L}_{1}-\mathcal{L}_{2}=
\partial_{\mu}X^{\mu}\xrightarrow{\textrm{MCP}}\mathcal{L}_{1}^{\ast}-\mathcal{L}_{2}^{\ast}
=D_{\mu}X^{\mu}.
\label{AM nas lagrangeanas}
\end{equation}
Therefore, if the equivalence class of Lagrangians $[\mathcal{L}]$ is required to be preserved
under the MCP, we need to have
\be
D_{\mu}X^{\mu}=\diverg_\tau (X),
\label{divs}
\ee
or, equivalently,
\be
\label{eq}
\pounds_X \, \tau = \left( D_{\mu}X^{\mu} \right)\tau.
\ee

The necessary and sufficient conditions to the existence of solutions for (\ref{divs})
can be obtained from the following proposition, whose (simple) proof is left to the Appendix.

\begin{prop} Let $\cal M$ be a differentiable manifold endowed
with a metric $g$ and a connection $\nabla$.
Let $\tau=f\tau_{0}$ be a volume form on $\cal M$.
Then, for every vector field $X$ on $\cal M$,
\[
D_{\mu}X^{\mu} = \diverg_\tau X + t(X) + v(X) - X(\ln f),
\]
where $t$ and $v$ are defined in eqs.~(\ref{1-forms}).
\label{prop divergencias}
\end{prop}

\subsection{ MCP in Affine manifolds}

In order to illustrate the previous results,
let us recall their application to the case of the metric-affine theory of gravity\cite{newhehl},
where spacetime is represented by a manifold endowed with a connection with possibly
nonzero curvature, torsion and nonmetricity.
We note that this contains, as a particular case, the Einstein-Cartan theory
of gravity, where the associated connection presents curvature and torsion but
is still metric-compatible ({\em i.e.}, $D_\kappa g_{\mu\nu}=0$).

For this case, Proposition~\ref{prop divergencias} implies that the
canonical volume form $\tau_0$ yields
\be
D_\mu X^\mu - \diverg_{\tau_0} X = (T_\mu + V_\mu) X^\mu.
\ee
It follows from our previous discussion that equivalent Lagrangian densities
of Special Relativity will be mapped, via MCP, to {\em nonequivalent} Lagrangian densities
in the corresponding non-flat spacetime. Therefore, since there is no
canonical representative of $[{\cal L}]$ in Special Relativity, the MCP turns out to
be essentially ill-defined in this context.

This difficulty can be circumvented with the help of Proposition~\ref{prop divergencias}
itself. The above incompatibility arises only when $D_\mu X^\mu \ne \diverg_{\tau} X$. However,
admitting a more general volume form $\tau = f\tau_0$ on $\cal M$, one can get
$D_\mu X^\mu = \diverg_\tau X$ by choosing $f$ such that
\be
t(\partial_\mu) + v(\partial_\mu) - \partial_\mu\ln f  =0,
\ee
which is equivalent to
\be
\label{cond1}
T_\mu + V_\mu =\partial_\mu \theta, \quad {\rm and\ } \tau = e^\theta \tau_0.
\ee
Therefore, the MCP is well defined (preserves equivalence classes
of Lagrangians) precisely when:
\begin{enumerate}
\item $t+v$ is an exact form, say $t+v=d\theta$, and
\item the volume element on $\cal M$ is given by $\tau = e^\theta \tau_0$.
\end{enumerate}
Again, since in the context of flat space (Special Relativity) there is no
canonical choice of representative of a given class of Lagrangians, our results
imply that the MCP is well defined only in the presence of the modified volume element
(\ref{mve}).

This is our main result: in order to preserve a given class $[{\cal L}]$
of Lagrangians under the MCP, one must have
$T_\mu + V_\mu =\partial_\mu \theta$ and, obligatorily, to use the modified
volume element (\ref{mve}). If the sum of traces  $T_\mu + V_\mu$ is not
a gradient, the MCP turns out to be ill-defined, since it produces
nonequivalent theories out of equivalent flat-space Lagrangians,
even at the classical level.

\subsection{ The $\lambda$-symmetry}

In an affine manifold, the Riemann tensor is invariant under the transformation
\be
\label{lambda}
\Gamma_{\mu\nu}{}^\kappa  \rightarrow \Gamma_{\mu\nu}{}^\kappa + \delta^\kappa_\nu\partial_\mu\lambda,
\ee
where $\lambda$ is an arbitrary function. This is the so-called $\lambda$-symmetry \cite{Einstein},
introduced by Einstein in his pioneering work on unified field theory, which
has been proved to be very useful in the analysis of affine theories of gravity\cite{Guendelman}.
In a $\lambda$-invariant theory, the function $\lambda$ can be properly chosen in order
to cancel some parts of the torsion tensor or/and of the non-metricity, simplifying
the overall analysis. Under (\ref{lambda}), the torsion tensor and $V_{\mu\nu\kappa}$
are changed as
\begin{eqnarray}
\label{lten}
T_{\mu\nu\kappa} &\rightarrow& T_{\mu\nu\kappa} + \left(g_{\kappa\nu}\partial_\mu\lambda - g_{\kappa\mu}\partial_\nu\lambda
\right), \nonumber \\
V_{\mu\nu\kappa} &\rightarrow& V_{\mu\nu\kappa} + \left(g_{\kappa\nu}\partial_\mu\lambda + g_{\kappa\mu}\partial_\nu\lambda
- g_{\mu\nu}\partial_\kappa\lambda\right),
\end{eqnarray}
implying to their traces  
\be
\label{ltrace}
T_\nu \rightarrow T_\nu - (n-1)\partial_\nu\lambda\quad {\rm and}\quad V_\nu \rightarrow V_\nu +n\partial_\nu\lambda.
\ee
One sees from (\ref{lten}) that a $\lambda$-transformation does not preserve the spaces of
symmetric nor metric-compatible connections.
However, the condition ensuring the consistency of MCP, namely that $T_\mu+V_\mu$ be a gradient,
is indeed preserved under (\ref{lambda}).
The Einstein-Hilbert action
is, obviously,   also preserved under (\ref{lambda}). Hence, 
provided that the matter content action is also $\lambda$-invariant,
the $\lambda$-transformation will map distinct solutions in a metric-affine theory of gravity
governed by the Einstein-Hilbert action, 
preserving all the isometries if $\lambda$ is properly chosen.

An interesting consequence arises from the transformation of
the modified volume element  under a $\lambda$-transformation,
\be
e^\theta\tau_0\rightarrow e^{\theta + \lambda} \tau_0.
\ee
Starting with, for instance, a Riemann-Cartan manifold with a gradient torsion
trace and an Einstein-Hilbert action with the modified volume element, 
it is possible to choose $\lambda=-\theta$ and to get an equivalent
theory, where the MCP is also well-defined, formulated in an affine manifold
(the connection is not metric-compatible anymore) but with the usual volume
element. In other words, in this new manifold, $D_\mu X^\mu =\diverg_{\tau_0} X$
for any vector field $X$.

\section{Final Remarks}

Before discussing possible applications of the results presented here to the
study of crystalline defects, let us recall how they reinforce the results
of \cite{Neutron} about
the inadequacy of the
propagating torsion model proposed in \cite{Saa1} to explain the available observational
data. The modified volume element (\ref{mve}) was introduced in the model of \cite{Saa1}
by means of exclusively geometrical arguments, and, hence, there is no
free parameter associated with the field $\theta$, which is responsible for the
propagating components of the torsion. Moreover, by employing the MCP, no extra parameters,
besides the Newtonian constant $G$, should arise in any gravitational model.
Thus, the only way of fitting the available observational data described in
\cite{Neutron} is to have $\theta$ constant, leading to the already mentioned conclusion
that the results of \cite{Neutron} would allow us to deduce that the
observational data and the very reasonable requirement of consistency of
the action formulation with the MCP {\em imply} that the trace of the torsion
tensor of spacetime should vanish.

Condensed matter is a vast subject but, to the best of our knowledge, the ideas
presented here were not yet applied to the study of crystalline defects,
which possess a geometrical description since the seminal works of
K. Kondo in the fifties (see, for references, \cite{Kroner}). The physics
of crystalline defects is a promising arena to text experimentally the ideas
of the last sections, with special emphasis to new effects that might be associated
with the modified volume element (\ref{mve}). The first conclusion about the
modified volume element in this context is that, in principle, it has no relation
to the {\em real} volume (area, in this case) of the polygon depicted in Fig. 1.
As it is discussed in \cite{Kroner}, contributions to the elementary area
coming from the Burgers vector are of higher order and should vanish
in the linear continuum limit. Incidentally, for $X^\mu$ and $Y^\mu$ orthogonal
in a Riemann-Cartan manifold ($D_\kappa g_{\mu\nu}=0$),
the area of the polygon of Fig.~1 is given, up to higher order terms, by
\be
A = |X||Y|\left(1-\frac{1}{2}\left(X^\mu + Y^\mu \right)T_\mu \right),
\ee
from where one sees that, despite of being unclosed, its area coincides
with the area of the parallelogram with sides $X^\mu$ and $Y^\mu$ if $T_\mu=0$.
However, the modified volume
element (\ref{mve}) cannot represent a physical area since any geometrical
property of the crystalline lattice should be invariant under a transformation
of the type $\theta\rightarrow\theta+\Lambda$, with constant $\Lambda$, and
the modified volume element, under this transformations, behaves as
$\tau\rightarrow e^\Lambda \tau$. This behavior, on the other hand, is allowed
in the minimal action formulation, since, as it was already noted, a scale transformation
of a Lagrangian  preserves the dynamics. The dynamics of defects, on the other
hand, are
governed by their free energy, which
can be expressed by means of a Lagrangian density related to
the Einstein-Hilbert Lagrangian\cite{kata,tartaglia}. The use of the modified
volume element in the action of the free energy would define a new dynamics
to the   torsion and metric tensors, analogous to that proposed
 in \cite{Saa1} in the context of
Einstein-Cartan theory of gravity.
An immediate consequence would
be the propagation of the trace of the torsion tensor and, consequently, of the
associated Burgers vector.

The free energy plays the role of the kinetic term in the geometrical action formulation
of defects. However, kinetic terms are not, a priori, specified by the MCP. Our analysis
seems more suitable to the study of fields/particles propagating on the crystalline lattice,
such as phonons, impurities and vacancies \cite{kata}.
If the perturbation energy associated with these objects is small enough, their
backreaction on the lattice may be ignored, and they can be considered {\em test} fields.
Phonons are nothing more than small elastic (sound) excitations in the crystal;
in the geometrical description, they correspond
to gauge independent linearized perturbations of the metric \cite{Solid}.
On the other hand, one can also consider impurities or vacancies moving on the
background medium  defined by the solid with crystalline defects.
The effect of the background on such test fields/particles can be gotten
from the MCP by demanding that the corresponding equations for
phonons, impurities and vacancies be obtained, from the equations for the
regular lattice, by changing the ordinary derivative to the covariant one and
the Euclidean metric to the metric associated with the background
medium.
With use of MCP and the modified volume element, testable predictions for
the propagation of phonons, impurities and vacancies could arise.
For instance, the modified volume element should affect the wavefunctions
and energy spectrum of impurities and vacancies\cite{kata} in the presence of
a crystalline defect. The occurrence of bound states due to dislocations\cite{edge}
can also be studied in the light of the present results.
One probably, however, needs to go beyond the eikonal approximation for the
test fields, since in this case the trajectory of the associated particle
is assumed to be a
geodesic curve\cite{kata}, a good approximation for high-frequency fields,
but with no contributions from the torsion tensor.
These points are now under investigation.

\acknowledgments

This work was supported by FAPESP and CNPq. The authors are grateful to
Dr. C. Miranda for helpful discussions about crystalline defects, and to
the Abdus Salam International Centre for Theoretical Physics, Trieste, Italy,
where this work was partly done, for the hospitality.

\appendix

\section*{Appendix}

Let us first discuss how the results of this paper relate to the earlier approach
of \cite{Saa1, Fiziev, Saa2, SaaCQG, Saa3}.
Unlike the approach presented here, the analysis of these works explicitly uses the
concept of covariant derivative of densities, which are defined as follows.
Let $h$ be a scalar density on $\cal M$. One can define the covariant derivative
of $h$ in two different ways:
\be
\label{usual}
D_{\mu} h=\partial_{\mu}h-\Gamma_{\mu\kappa}{}^{\kappa}h
\ee
or
\be
\label{transp}
\tilde{D}_{\mu} h=\partial_{\mu}h-\Gamma_{\kappa\mu}{}^{\kappa}h.
\ee
The definition (\ref{usual}) can be called the {\em usual} covariant derivative,
since, as we show in the following,
\be
\label{lieid}
\nabla_{X}(h \; dx^{1}\wedge\cdots\wedge dx^{n})=[X^{\mu}D_{\mu}h] \;
dx^{1}\wedge\cdots\wedge dx^{n}.
\ee
The covariant derivative (\ref{transp}), used also, for instance, in
\cite{Lovelock}, is the so-called transposed
derivative\cite{Fiziev}, since it can be interpreted as the covariant
derivative with respect to the transposed connection $\tilde{\Gamma}_{\mu\nu}{}^{\kappa}=
\Gamma_{\nu\mu}{}^{\kappa}$
(see, in this context, the volume-preserving connection introduced in
\cite{newhehl}).
Note that, according to eq.~(\ref{torsion}),
$\tilde{D}_{\mu} h= {D}_{\mu} h $ provided that the trace of the torsion
tensor vanishes so that, in this case, no ambiguity concerning the
definition of the covariant derivative of densities arises.
The covariant derivative (\ref{transp}) is
closer to the {\em Lie derivative} on $\cal M$, since it satisfies, as we will see,
\be
\pounds _{X}(h \; dx^{1}\wedge\cdots\wedge dx^{n})=[(D_{\mu}X^{\mu}) \; h+
X^{\mu}(\tilde{D}_{\mu} h)]dx^{1}\wedge\cdots\wedge dx^{n}.
\label{lie}
\ee

The mathematical property that distinguishes the volume element (\ref{mve})
in \cite{Saa3} is that, whereas the canonical volume element (\ref{canonical})
is covariantly constant with respect to the usual derivative,
\be
 {D}_\mu  \sqrt{g} = 0,
\ee
the volume element (\ref{mve}) is covariantly constant with respect to the transposed derivative,
\be
\tilde{D}_\mu (e^\theta \sqrt{g}) = 0,
\label{Dmodvol}
\ee
provided that
\be
T_\mu+V_\mu=\partial_\mu\theta.
\ee
By comparing with eq.~(\ref{lie}), we see that (\ref{eq}) holds if and only if
one chooses a volume form $\tau=f\tau_0$ such that
\be
\label{last}
\tilde{D}_\mu (f\sqrt{g})=0,
\ee
which has a solution only if $T_\mu+V_\mu=\partial_\mu\theta$ and $f=e^\theta$ \cite{Saa3}.
This shows how the earlier approach of \cite{Saa1, Fiziev, Saa2, SaaCQG, Saa3} relates
to the present work, which, as discussed in the previous sections, is based on weaker physical assumptions.

We end by providing proofs of identities (\ref{lieid}) and (\ref{lie}),
and of Proposition~\ref{prop divergencias}.
Identity (\ref{lieid}) can be proved as follows:
\begin{eqnarray}
\nabla_{X}(hdx^{1}\wedge\cdots\wedge dx^{n})&=&
 (\nabla_{X}h)~dx^{1}\wedge\cdots\wedge dx^{n}
 + h\textstyle\sum\nolimits_{\kappa}dx^{1}\wedge\cdots\wedge\nabla_{X}dx^{\kappa}\wedge\cdots\wedge
 dx^{n} \nonumber \\
& = & X(h)\,dx^{1}\wedge\cdots\wedge dx^{n}+h\textstyle\sum\nolimits_{\kappa}dx^{1}\wedge\cdots
\wedge(-X^{\mu}\Gamma_{\mu\nu}{}^{\kappa}dx^{\nu})\wedge\cdots\wedge dx^{n}
 \nonumber \\
& = &X^{\mu}(\partial_{\mu}h-\Gamma_{\mu\kappa}{}^{\kappa})\,dx^{1}\wedge\cdots\wedge dx^{n}.
\end{eqnarray}
As for (\ref{lie}), since $\pounds _{X}$ and $d$ commute, we have
\be
\pounds _{X}dx^{\kappa}=
d\pounds _{X}x^{\kappa}=dX^{\kappa}=\partial_{\nu}X^{\kappa}dx^{\nu},
\ee
which leads to
\begin{eqnarray}
\pounds _{X}(hdx^{1}\wedge\cdots\wedge dx^{n}) &=& \pounds _{X}(h)~dx^{1}\wedge\cdots\wedge dx^{n}+
h\textstyle\sum\nolimits_{\kappa}dx^{1}\wedge\cdots\wedge\pounds _{X}dx^{\kappa}\wedge\cdots\wedge dx^{n} \\
&=& [X(h)+h\partial_{\kappa}X^{\kappa}]\,dx^{1}\wedge\cdots\wedge dx^{n} \label{eqq}\\
&=& [hD_{\mu}X^{\mu}+X^{\mu}(\partial_{\mu}h-h\Gamma_{\kappa\mu}{}^{\kappa})] \, dx^1 \wedge\cdots\wedge dx^n.
\end{eqnarray}

Finally, Proposition~\ref{prop divergencias} can be proved as follows.
We first notice that
\begin{eqnarray}
D_\kappa X^\kappa &=& \partial_\kappa X^\kappa + \Gamma_{\kappa\mu}{}^\kappa X^\mu \\
&=& \partial_\kappa X^\kappa + ( \genfrac{\{}{\}}{0pt}{1}{\kappa}{\kappa\mu} + K_{\kappa\mu}{}^\kappa + V_{\kappa\mu}{}^\kappa ) X^\mu \\
&=& \partial_\kappa X^\kappa + \left( \partial_\kappa\ln\sqrt g + T_\kappa + V_\kappa \right) X^\kappa .
\end{eqnarray}
On the other hand, it follows from eq.~(\ref{eqq}) that
\be
\pounds_X \left( f \vol \right) =\left( X^\kappa\partial_\kappa\ln(f\sqrt g) + \partial_\kappa X^\kappa \right) \tau.
\ee
Thus, from definition (\ref{def diverg}),
\be
\diverg_\tau X = X^\kappa\partial_\kappa\ln(f\sqrt g) + \partial_\kappa X^\kappa,
\ee
which finally yields
\be
D_\kappa X^\kappa =\diverg_\tau X + \left( T_\kappa + V_\kappa -\partial_\kappa\ln f \right) X^\kappa.
\ee

\end{document}